\documentclass[12pt,preprint,a4paper]{aastex}

\usepackage{graphics,epsf}
\usepackage{amsmath}                
\usepackage{amsfonts}               
\usepackage{amssymb}                
\usepackage{epsfig}                 

\def \yr{~\rm{yr}}

\def \astrobj#1{#1}

\begin{document}

\title{SPINNING-UP THE ENVELOPE BEFORE ENTERING A COMMON ENVELOPE PHASE}

\author{Ealeal Bear\altaffilmark{1} and Noam Soker\altaffilmark{1}}

\altaffiltext{1}{Department of Physics, Technion$-$Israel
Institute of Technology, Haifa 32000 Israel;
ealealbh@gmail.com; soker@physics.technion.ac.il.}
\setlength{\columnsep}{1cm} \small

\begin{abstract}
We calculate the orbital evolution of binary systems where the
primary star is an evolved red giant branch (RGB) star, while the
secondary star is a low-mass main sequence (MS) star or a brown
dwarf. The evolution starts with a tidal interaction causes the
secondary to spiral-in. Than either a common envelope (CE) is
formed in a very short time, or alternatively the system reaches
synchronization and the spiraling-in process substantially slows
down. Some of the latter systems later enter a CE phase. We find
that for a large range of system parameters, binary systems reach
stable synchronized orbits before the onset of a CE phase. Such
stable synchronized orbits allow the RGB star to lose mass prior
to the onset of the CE phase. Even after the secondary enters the
giant envelope, the rotational velocity is high enough to cause an
enhanced mass-loss rate. Our results imply that it is crucial to
include the pre-CE evolution when studying the outcome of the CE
phase. We find that many more systems survive the CE phase than
would be the case if these preceding spin-up and mass-loss phases
had not been taken into account. Although we have made the
calculations for RGB stars, the results have implications for
other evolved stars that interact with close companions.
\end{abstract}


\section{Introduction}
\label{sec:intro}

Extreme horizontal branch (EHB) stars are post red giant branch
(RGB) stars that burn helium in their core and have a very
low-mass envelope. Spectroscopically they are classified as sdB
(or sdO) stars, although some sdB and sdO stars might be post
asymptotic giant branch (AGB) stars. In what follows we will refer
by sdB stars only to EHB stars, and include under the name sdB
stars also the sdO stars. Many of the sdB stars in the field
(i.e., not in globular clusters) are in binary systems (Maxted et
al. 2001a,b), some with a close and very faint M type main
sequence (MS) companion, e.g., \astrobj{HS~$0705+6700$} (Drechsel
et al 2001), \astrobj{PG~$1336-018$} (Kilkenny et al. 1998), and
\astrobj{HW~Vir} (Menzies \& Marang 1986; Wood et al. 1993).

The orbital and physical parameters of the three systems
\astrobj{HS~$0705+6700$}, \astrobj{PG~$1336-018$}, and HW~Vir, are very similar (Drechsel et al 2001).
In \astrobj{HS~$0705+6700$}, for example, the sdB and secondary stellar masses and radii are
$M_1=0.483 M_\odot$, $M_2=0.134 M_\odot$, $R_1=0.230 R_\odot$, and $R_2=0.186 R_\odot$, respectively,
while the orbital separation and period are $0.81 R_\odot$, and $2^{\rm h}18^{\rm m}$, respectively.
It turns out that the secondary star almost fills its roche lobe (Drechsel et al 2001).
Typically these systems have $q \equiv M_2/M_1 \la 0.3$.
Detection is biased toward close and more massive companions.
This suggests that many more systems exist with even lower values of $q$ (Davis et al. 2009).
Geier et al. (2009), for example, announced recently the detection of a $8 - 23 M_J$,
where $M_J$ is Jupiter mass, substellar companion orbiting an sdB star at an orbital
separation of $5-6.1 R_\odot$.

These systems must have gone through a common envelope (CE) phase
(e.g.,Livio \& Soker 1988; Drechsel et al. 2001; Han et al. 2007
and Han 2008 for recent papers with more references). Different
aspects regarding the formation of sdB stars and the evolution of
RGB stars through a CE phase have been discussed in the literature
for more than three decades (e.g., Pustynski \& Pustylnik 2006;
Soker 2006; Nordhaus \& Blackman 2006; Nordhaus et al 2007; Livio
\& Soker 2002; Dewi \& Tauris 2000; D`Cruz et al 1996; Alexander
1976; Paczynski 1971; Han et al. 2002, 2003, 2007; Han 2008;
Zuckerman et al 2008; Melis et al 2009). However, some questions
remain open. (1) Does the companion transfer part, or all, of the
mass it accreted during the CE phase back to the primary at the
end of the CE phase? (2) What is the role played by very low-mass
MS companions, considering that even substellar objects (brown
dwarfs and planets) can form sdB stars, as suggested by one of us
in a series of papers (Soker 1998, Soker \& Harpaz 2000, 2007),
and confirmed observationally by Geier et al. (2009)?

In the present paper we examine how tidal interaction before the
formation of the CE might increase the likelihood of a low-mass MS
companion to survive the CE phase. In section \ref{sec:CE1} we
explain the basic mechanism and present simple solutions ignoring
the evolution of the primary star along the RGB (but mass-loss is
included). In section \ref{sec:Momentum2} we also consider the
increase in the stellar core mass and stellar radius. We summarize
our main results in section \ref{sec:summary}.

\section{Basic characteristics of the evolution}
\label{sec:CE1}

The onset of the CE, and the ejection of the envelope before and
after the onset of the CE, have received a great deal of
attention over the years (e.g., Soker et al 1984; Tauris \& Dewi
2001; Han et al. 2002; De Marco et al. 2003; Soker 2004; Hu et al.
2007; Nelemans \& Tout 2005; Beer et al. 2007; Politano \& Weiler
2007; Webbink 2008; Davis et al. 2009). Our goal is to study in
more detail the evolution of binary systems in a stage prior to the
onset of the CE phase, and in particular systems that have reached
synchronization;
the synchronization is between the orbital period and the primary rotation period.

We start our calculation when tidal interaction becomes important.
For the binary systems we study, where the primary is an RGB star
and the secondary is a low-mass main sequence (MS) star or a brown
dwarf, tidal interaction becomes important when the giant swells
to a radius of $R_g \sim 0.2 a$, where $a$ is the orbital
separation (Soker 1998). The primary radius increases along the
RGB as the core mass increases. The primary and the secondary
initial masses range in the binary systems we study are
$0.8M_{\odot} \le M_1 \le 2.2M_{\odot}$, and $0.015M_{\odot} \le
M_2 \le 0.2M_{\odot}$, respectively.

When a binary system starts its evolution it is not synchronized,
and therefore tidal interaction will lead to a fast spiraling-in
process, i.e., the orbital separation decreases. The binary system
can then either reach a synchronization or stays asynchronous. In
the asynchronous systems tidal interaction will lead to a fast
spiraling-in process all the way to a rapid formation of a CE. In
systems where synchronization is reached the orbital evolution
substantially slows down (unless a Darwin instability occurs as
soon as synchronization is achieved; see below), as its rate is
now controlled by mass-loss and the evolution of the primary star;
the orbital separation can decrease or increase.

When the orbital separation increases faster than the RGB stellar
radius, the evolution is terminated when all the primary envelope
is lost, or a core helium flash occurs; no CE phase occurs. If the
orbital separation decreases relative to the RGB stellar radius,
the evolution ends when at least one of the following conditions
occur:
\newline
(1) When the mass of the primary envelope has been totally lost.
No CE phase occurs.
\newline
(2) When the core experiences a core helium flash.
Here we take this to occur when the core mass reaches $0.48 M_\odot$.
No CE phase occurs.
\newline
(3) Darwin Instability (see below) is reached, the secondary falls into the
primary envelope.
\newline
(4) The RGB stelar radius increases to the point where it engulfs the
secondary, namely, $R_g>a$.
If conditions (3) or (4) are met, the system enters a CE phase after
some fraction, even a large fraction,  of the primary envelope has been lost.

\begin{figure}
\includegraphics[scale=0.6]{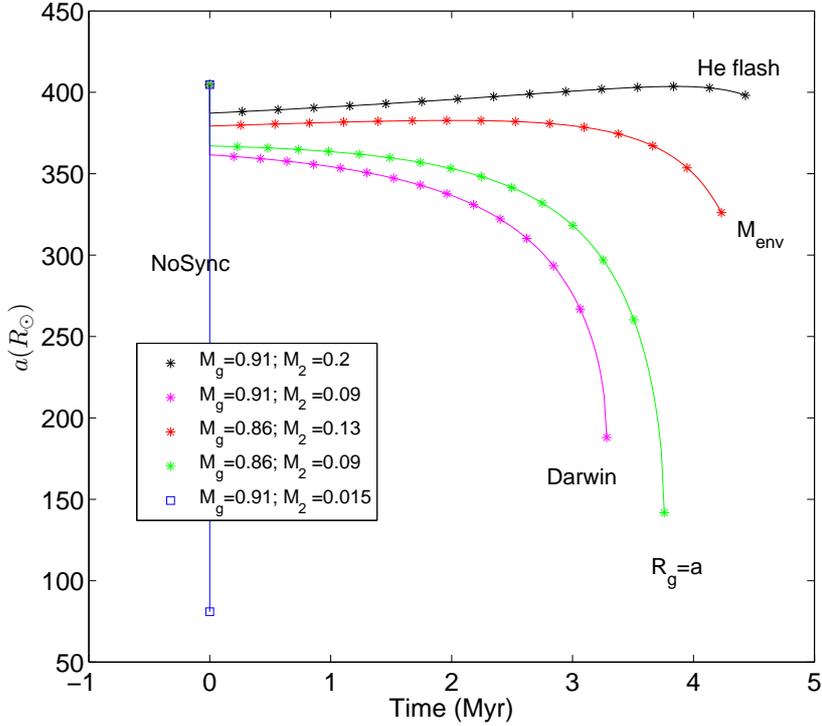}
\caption{ Orbital separation, in solar radii for different
binaries as indicated, vs. time. The termination of the evolution
is indicated. NoSync: No stable synchronized orbit is reached, and
a rapid CE phase occurs; $M_{\rm env}:$ the RGB has lost its
envelope; $R_g=a:$ the RGB stellar radius increased to the point
where it swallowed the secondary; Darwin: the Darwin instability
caused the formation of a common envelope; He flash: the core mass
reached a mass of $0.48 M_\odot$ and a He core flash was assumed.
The initial RGB stellar mass $M_g$ and secondary mass $M_2$ for
each case are indicated in the legend. In all cases the mass-loss
parameter in equation (\ref{dlx5}) is $\eta_R=3 \times 10^{-13}$,
the initial core mass is $M_{c}(0)=0.4 M_\odot$, the orbital
separation prior to tidal interaction is $a_0=5R_g(0)=405
R_\odot$, and the calculations include the evolution of the
primary star as explained in section \ref{sec:Momentum2}.}
\label{fig:sdbf1}
\end{figure}

The different types of behavior are illustrated in Fig.
\ref{fig:sdbf1}. This figure includes the evolution of the primary
star, and the detail of the calculations will be presented in
section \ref{sec:Momentum2}. Here we are only interested in the
general outcome. The orbital separation (in units of $R_\odot$)
versus time is presented for several cases, as indicated in the
figure. When the secondary mass is too low (exact value depends on
the RGB stellar properties) the secondary star does not have
enough orbital angular momentum to bring the primary rotation in
synchronism with the orbital period. Tidal interaction rapidly
cause the secondary to spiral-in and form a CE. We do not
calculate the tidal time scale, as it is very short.

When the secondary mass is larger than some critical value for the
same RGB star, a synchronization is achieved (in a short time)
before the orbital separation decreases to the RGB radius. (As we
do not treat the initial process of reaching synchronization, the
calculation seems to start from a radius of smaller that the
primordial orbital separation.) Now the system evolves on a
mass-loss time scale. The primary star loses mass, an effect that
tends to increase orbital separation. However, the wind carries
angular momentum. To maintain synchronization the secondary
transfers, via tidal forces, orbital angular momentum to the
envelope, and the orbit shrinks. Tidal effects might overcomes the
mass-loss effect. Eventually, a point is reached where the
secondary enters the envelope, either due to the Darwin
instability, or by the swelling RGB envelope that increases to
$R_g=a$. By that time the envelope mass is lower than its initial
value. Calculations that do not consider the possibility of pre-CE
mass-loss might attribute the entire mass-loss to the CE phase.
Such calculations overestimate the efficiency by which the
secondary expels the envelope during the CE phase. This
overestimate can lead to a search for an extra energy source that
is actually not required.

When the secondary mass is larger even (for the same RGB star),
the secondary orbit shrinks by a small amount before it brings the
primary to synchronization. As the orbital separation is large,
the effect of mass-loss overcomes the effect of angular momentum
transfer by tidal interaction and the orbital separation
increases.

To demonstrate the basic characteristics of the evolution, we
start by neglecting the evolution of the RGB star. Namely, we
assume the core mass and RGB stellar radius to be constant, as in
Soker (2002). As a single star reaches this radius its rotation
angular velocity $\omega$ will be very small; practically zero.
When synchronization is achieved, it is done on a short time, and
we neglect angular momentum loss in the wind, such that the source
of most of the RGB angular momentum $J_{\rm env}=I_{\rm env}
\omega $, is the change in the orbital angular momentum $\Delta
J_{\rm orb}$, where $I_{\rm env}$ is the envelope's moment of
inertia. Namely,
\begin{equation}
I_{\rm env} \omega_{\rm syn}= \Delta J_{\rm orb},
\label{dlx1}
\end{equation}
where $\omega_{\rm syn}$ is the angular velocity of the primary envelope when
synchronization is achieved for the first time, and
\begin{equation}
\Delta J_{\rm orb} =
\mu (G M)^{0.5}  \left( a_0^{0.5}-a_{\rm syn}^{0.5} \right),
\label{dlx2}
\end{equation}
where $a_0$ is the orbital separation prior to synchronization
(when tidal interaction becomes important), and is set here to
$a_0=5R_g(0)$, and $a_{\rm syn}$ is the orbital separation when
synchronization is achieved for the first time. As usual, $M
\equiv M_g + M_2$ and $\mu \equiv M_g M_2 /M$ where $M_g=M_1$ is
the mass of the giant. Synchronization implies the equality
\begin{equation}
\omega_{\rm syn}=\left[ G M_{g}(a_{\rm syn})^{-3} \right] ^{0.5},
\label{dlx3}
\end{equation}
for $M_2 \ll M_g$.

Due to its small size, the core's angular momentum is negligible.
The angular momentum carried by the wind must be considered after
synchronization is achieved because the evolution slows down, and
the primary's mass-loss rate on the RGB is significant over a long
time. Actually, under the assumption that the primarys core does
not evolve, the evolution is controlled by mass-loss. Conservation
of angular momentum determines the evolution
\begin{equation}
\dot{J}_{\rm orb}+ \dot{J}_{\rm env}+\dot{J}_{\rm wind}=0.
\label{dlx4}
\end{equation}

Using the evaluation of equation (\ref{dlx4}) as given by Soker
(2002), we solved for the orbital evolution as function of
mass-loss for an RGB star with constant core mass of $ M_c= 0.4M_{
\odot}$ and radius of $R_g(0)= 81R_{\odot}$. Tidal interaction is
assumed to start when the stellar radius grows to $R_g=0.2a_0$,
hence here we set $a_0= 5R_g(0) = 405 R_{ \odot}$, and a constant
core mass of $ M_c= 0.4M_{ \odot}$ (for details see Soker 2002).
The RGB radius is fixed at $R_g(0)= 81R_{\odot}$. When the
evolution of the primary is not considered, the orbital separation
is a function of the envelope mass. For that case there is an
analytical solution to determined if a stable synchronization
orbit  is achieved.

In some cases when synchronization is achieved, the system is already unstable
to the Darwin instability.
This instability sets in when a small decrease in the orbital radius does not
transfer enough angular momentum to the spinning primary to maintain a synchronization.
The condition for stability is $I_{\rm orb} > 3I_{\rm env}$, where
$I_{\rm orb}=M_gM_2/(M_g+M_2)a^2 $ is the orbital moment of inertia, and
$I_{\rm env}=\eta M_{\rm env} R_g^2$ for the envelope's moment of inertia.
For RGB stars $\eta \simeq 0.2$.
Taking again $M_2 \ll M_g$, the instability sets in when
\begin{equation}
M_2 a^2 =3 \eta M_{\rm env} R_g^2.
\label{eq:darwin1}
\end{equation}

\begin{figure}
\includegraphics[scale=0.6]{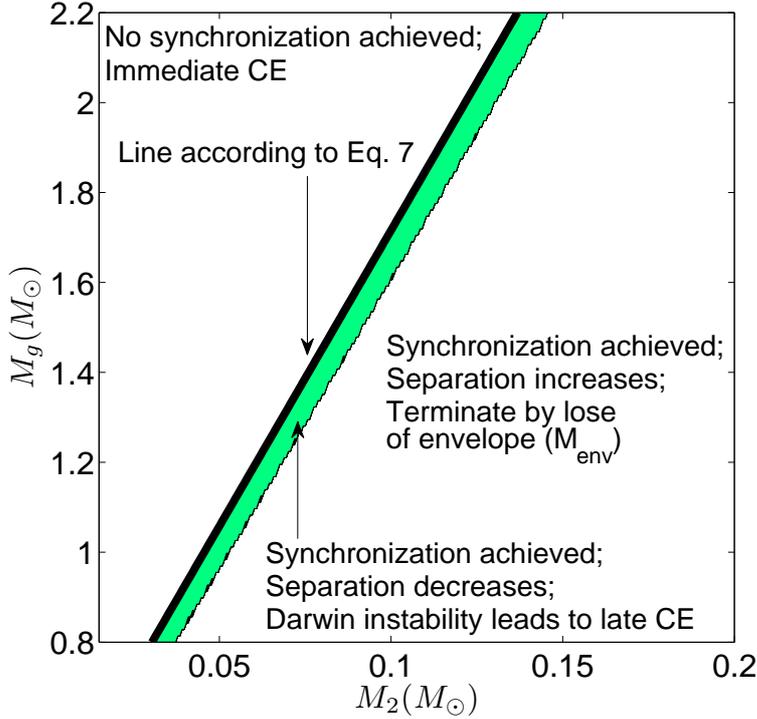}
\caption{ $M_g$  vs. $M_2$  for binaries with a constant RGB core
mass $M_c$, and hence constant radius $R_g$ (RGB evolution is not
considered). The upper-left trapeze represents binaries that did
not achieve synchronization, but rather formed an immediate CE.
The lower-right trapeze represents binaries that achieved
synchronization, but never formed a CE. In these cases the
envelope was depleted, mainly due to the RGB wind. The thick strip
in the $M_2/M_g$ plane is for binaries that achieved
synchronization before they entered a late CE. The straight line
represents equation (\ref{eq:darwin3}). In all cases here the
mass-loss parameter in equation (\ref{dlx5}) is $\eta_R=3 \times
10^{-13}$, the core mass is $M_{c}=0.4 M_\odot$, and the orbital
separation prior to onset of tidal interaction (primordial
separation) is $a_0=5R_g(0)=405 R_\odot$.} \label{fig:sdbf2}
\end{figure}

Equations (\ref{dlx1}), (\ref{dlx2}), (\ref{dlx3}) and (\ref{eq:darwin1}) can be combined to
yield the condition that a stable synchronized orbit is achieved, that for
$M_2 \ll M_g$ can be approximated as
\begin{equation}
M_2 > \frac{4^4 \eta}{3^3} \left( \frac{a_0}{R_g} \right)^{-2} M_{\rm env}.
\label{eq:darwin2}
\end{equation}
Taking $a_0=5R_g$ as before, $\eta=0.2$, and $M_{\rm env}=
M_g-0.4$, for a core mass of $M_c = 0.4 M_\odot$ (masses in solar
units if not indicated otherwise), in equation (\ref{eq:darwin2}) gives
\begin{equation}
M_2 > 0.076M_{\rm env} = 0.076 (M_g -0.4 M_\odot).
\label{eq:darwin3}
\end{equation}
This represents the straight line in the $M_g - M_2$ plane, drawn
as straight blue line in Fig.~\ref{fig:sdbf2}.

In Fig.~\ref{fig:sdbf2} we present by a wide strip the region in
the $M_g - M_2$ plane where the systems reach synchronization
before the onset of a CE. Above this strip the companion spiral in
into the envelope before a synchronization is achieved. Below that
strip, the secondary avoids the CE. As found already by Soker
(2002), this strip is very narrow. The line according to equation
(\ref{eq:darwin3}) is also drawn in the figure, just on the
upper-left boarder of the strip. To better explore the possibility
of significant mass-loss prior to the onset of a common envelope,
we turn to include the evolution of the primary.

\section{Calculations with an evolving primary star}
\label{sec:Momentum2}

\subsection{Basic equations}
\label{subsec:equaitons}

When the evolution of the primary is not considered, the orbital
separation is a function solely of the envelope mass, as the core
mass and radius are constant. In that case the dependence on time
comes from the mass-loss rate. In this section the increase in
core mass and its influence on the primary RGB star are taken into
consideration. The mass-loss from the envelope is calculated by
using the Reimers equation (Reimers 1975; see, e.g., Catelan 2009,
Schoder \& Cuntz 2007, Suzuki 2007, and Meszaros et al. 2009 for
recent usages and discussions of that formula)
\begin{equation}
\dot{M_g}=\eta_R L_{\odot} R_{\odot}/ M_{\odot}
\label{dlx5}
\end{equation}
Although this formula is commonly used to estimate the mass-loss
from the giant, most researchers agree that it does not cover all
the physical mechanism and characteristics. For example, we know
from the different distributions of HB stars in different globular
clusters that other effects, such as metallicity, influence the
wind ejection process (e.g., D'Cruz et al 1996, 2000; Whitney et
al. 1998; Catelan 2000). The value of $\eta_R$ has been discussed
at length in the literature. Since it was shown that the Reimers
formula underestimates the wind ejected from some giants (Schroder
\& Kuntz 2005, Suzuki 2007), we will explore the results for two
values: $\eta_R =3 \times 10^{-14}$ (normal), and  $\eta_R =3
\times 10^{-13}$ (high).

In this section we consider the nuclear evolution of the star.
Nuclear burning increases the core mass, implying increase in luminosity and
RGB radius.
For the growth rate of the RGB core we take an expression given by Iben \& Tutukov (1984) and Nelemans \& Tauris (1998),
based on results of Mengel et al (1979)
\begin{equation}
\dot{M}_{c} = 10^{-5.36} \left( \frac {M_c}{M_\odot} \right)^{6.6} M_\odot \yr^{-1}.
\label{dlx10}
\end{equation}
The RGB stellar radius is calculated according to the relation proposed by Iben \& Tutkov (1984)
\begin{equation}
 R_{g}=10^{3.5}\left( \frac {M_c}{M_\odot} \right)^4  R_\odot
 \label{dlx12}
\end{equation}

The rate of angular momentum carried by the wind (defined
positively) in a synchronized system (primarys rotation
synchronized with orbital motion) is given by (Soker 2002)
\begin{equation}
\dot{J}_{\rm wind}=\left[ \left( \frac{M_2}{M_g+M_2} \right)^2 a^2+
\beta R_g^2 \right] \omega (-\dot{M_g}).
\label{dlx6}
\end{equation}
The first term is the angular momentum due to the orbital motion
of the primary around the system's center of mass, while the
second term is due to the rotation (spin) of the primary. Here
$-\dot M_g = \dot M_w >0$ is the mass-loss rate in the wind. We
neglect accretion by the secondary. The rate of change in the
orbital angular momentum $J_{\rm orb}=\mu (GM a)^{1/2}$, where
$\mu =M_gM_2/M$ is the reduced mass, and $M=M_g + M_2$ is the
total mass of the binary system, is given by
\begin{equation}
\frac{ \dot{J}_{\rm orb}}{{J}_{\rm orb}}=
\frac{ \dot M_g}{M_g}-0.5\frac{ \dot M_g}{M_g+M_2}+0.5\frac{\dot a}{a}.
\label{dlx7}
\end{equation}
 The rate of change of the envelope angular momentum $J_{\rm env}=I_{\rm env} \omega$ is
$\dot J_{\rm env}=\dot I_{\rm env} \omega + \dot \omega I_{\rm env}$.
Substituting $I_{\rm env}=\eta M_{\rm env} R_{g}^{2}$, the synchronization
condition $\omega = \sqrt{G(M_g+M_2)/a^3}$, and assuming that $\eta$ does not vary along
the RGB, gives after some rearrangement
\begin{equation}
\frac {\dot J_{\rm env}}{J_{\rm env}}=
\frac{\dot M_{\rm env}} {M_{\rm env}}
+2 \frac{\dot R_g}{R_g}
+0.5 \frac{ \dot M_g} { M_g+M_2 }
-1.5 \frac {\dot a}{a}.
\label{dlx8}
\end{equation}

Substituting Eqs. (\ref{dlx6})- (\ref{dlx8}) into Eq. (\ref{dlx4})
gives ($M=M_g+M_2$):
\begin{eqnarray}
\frac{\dot a}{a} \left[ 1-3\eta M_{\rm env} \frac{M}{M_g M_2} \left( \frac{R_g}{a} \right)^{2} \right]
=
-\frac{\dot M_g}{M}
\left[ 1-2\beta \frac{M^2}{M_g M_2} \left(\frac{R_g}{a} \right)^{2} \right]
\nonumber \\
-2 \eta \frac{ \dot M_{\rm env}M}{M_g M_2} \left( \frac{R_g}{a} \right)^{2}
- \eta \frac{M_{\rm env} \dot M_g}{M_g M_2} \left( \frac{R_g}{a} \right)^{2}
- 4 \eta M_{\rm env}\frac{\dot R_g}{R_g}  \frac{M}{M_g M_2} \left( \frac{R_g}{a} \right)^{2} .
\label{dlx9}
\end{eqnarray}

\subsection{Solution procedure}
\label{subsec:proceedure}

The solution was obtained under the following assumptions and by
performing the following  steps.
\newline
(1) We chose as initial conditions: Core mass $M_c$. This fixed
the RGB radius by equation (\ref{dlx12}).
 RGB stellar mass $M_g$; companion mass $M_2$; orbital separation $a$ with a circular orbit;
 coefficient $\eta_R$ of the mass-loss formula (\ref{dlx5}).
 For the RGB envelope's moment of inertia we take $\eta=0.2$, and for the angular momentum loss
 we take $\beta=2/3$ in equation (\ref{dlx6}), as appropriate for a uniform (spherical)
 mass-loss geometry from the surface of the RGB star.
 \newline
(2) The tidal interaction depends very strongly on the ratio of the stellar
radius to orbital separation $(R_g/a)$.
We therefore assume, as in section \ref{sec:CE1}, that tidal interaction
is negligible, until the RGB stellar radius is large enough, when a strong
tidal interaction occurs. We take this ratio to be $R_g=0.2a$.
\newline
(3) As in section \ref{sec:CE1}, the onset of the tidal interaction results
in one of three possibilities.
 ($i$) The orbital angular momentum is too low, and no synchronization
 is achieved. The companion spirals-in to the RGB envelope.
($ii$) Synchronization between the orbital motion and the RGB rotation is reached,
but the system at that stage is unstable to the Darwin instability (see eq. \ref{eq:darwin1}).
Here again, the companion spirals-in within a very short time to form a CE.
($iii$) Synchronization is achieved and the system is stable.
This evolutionary rout is further evolved, as described below.
\newline
(4) After synchronization is achieved in a stable system, we start to evolve the
primary star. The growth of the core mass is according to the rate of hydrogen
nuclear burning that forms helium, as given in equation (\ref{dlx10}).
This mass is removed from the envelope.
It changes the moment of inertia of the envelope, but not its angular momentum.
\newline
(5) Simultaneously with the growth of the core we consider the
wind, by removing mass and angular momentum from the RGB star,
according to equations (\ref{dlx5}) and (\ref{dlx6}),
respectively.
\newline
(6) As the wind caries angular momentum, and the growth of the core changes the
RGB  moment of inertia, in each time steps we solve equation (\ref{dlx9}) to determine
a new orbital separation $a$.
\newline
(7) In each time step we check whether one of the following occurs.
($i$) The RGB envelope has been completely depleted. We register the orbital separation.
($ii$) The RGB core has reached a mass of $M_c=0.48 M_\odot$, when a core helium flash
is assumed to occur.
We register the orbital separation.
($iii$) The Darwin instability occurs. In this case a late CE is formed. We register the
RGB envelope rotation velocity.
($iv$) The increasing radius of the RGB reached the orbital separation $R_g=a$.
In this case a late CE occurs.

\subsection{Results}
\label{subsec:results}
\subsubsection{Evolution of orbital separation and envelope mass}
\label{subsec:orbit}

\begin{figure}
\includegraphics[scale=0.6]{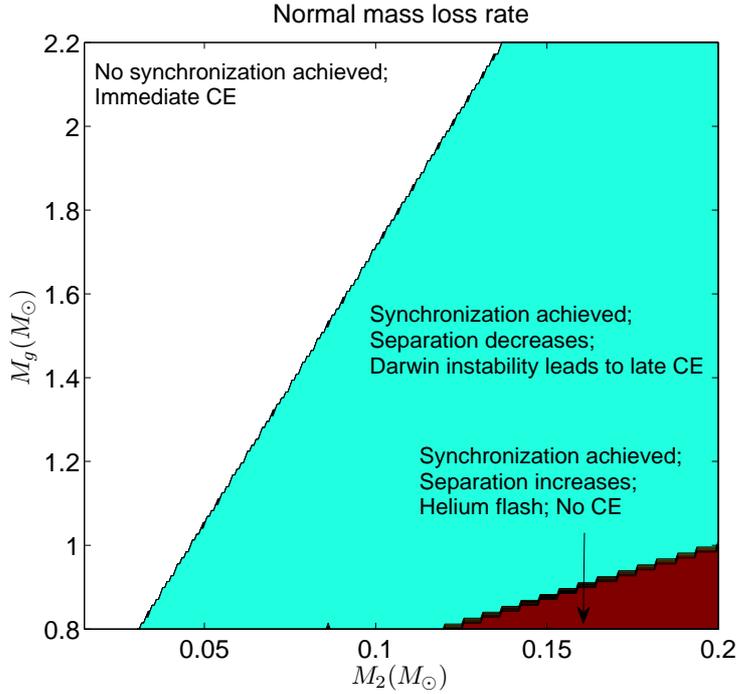}
\caption{ $M_g$ vs. $ M_2$ representing binary systems that
reached or did not reach synchronization. Our normal mass-loss
rate of $\eta_R=3 \times 10^{-14}$ was used. The initial core mass
is $M_{c}(0)=0.4 M_\odot$, and the initial (primordial) orbital
separation prior to tidal interaction is $a_0=5R_g(0)=405
R_\odot$. Calculation is terminated as indicated, when one of the
following occurs. (1) The Darwin instability brings the system to
a CE phase (marked `Darwin Instability'); (2) The RGB stellar
radius exceeds the orbital separation ($R_g = a$; does not happen
in this case); (3) A total depletion of the RGB envelope occurs
($M_{\rm env}$; does not occur in this case due to a relatively
low-mass loss rate as indicated by $\eta_R$). (4) The core mass
reaches $0.48 M_\odot$ and assumed to go through a core Helium
flash ('Helium flash').} \label{fig:sdbf3}
 \end{figure}

\begin{figure}
\includegraphics[scale=0.6]{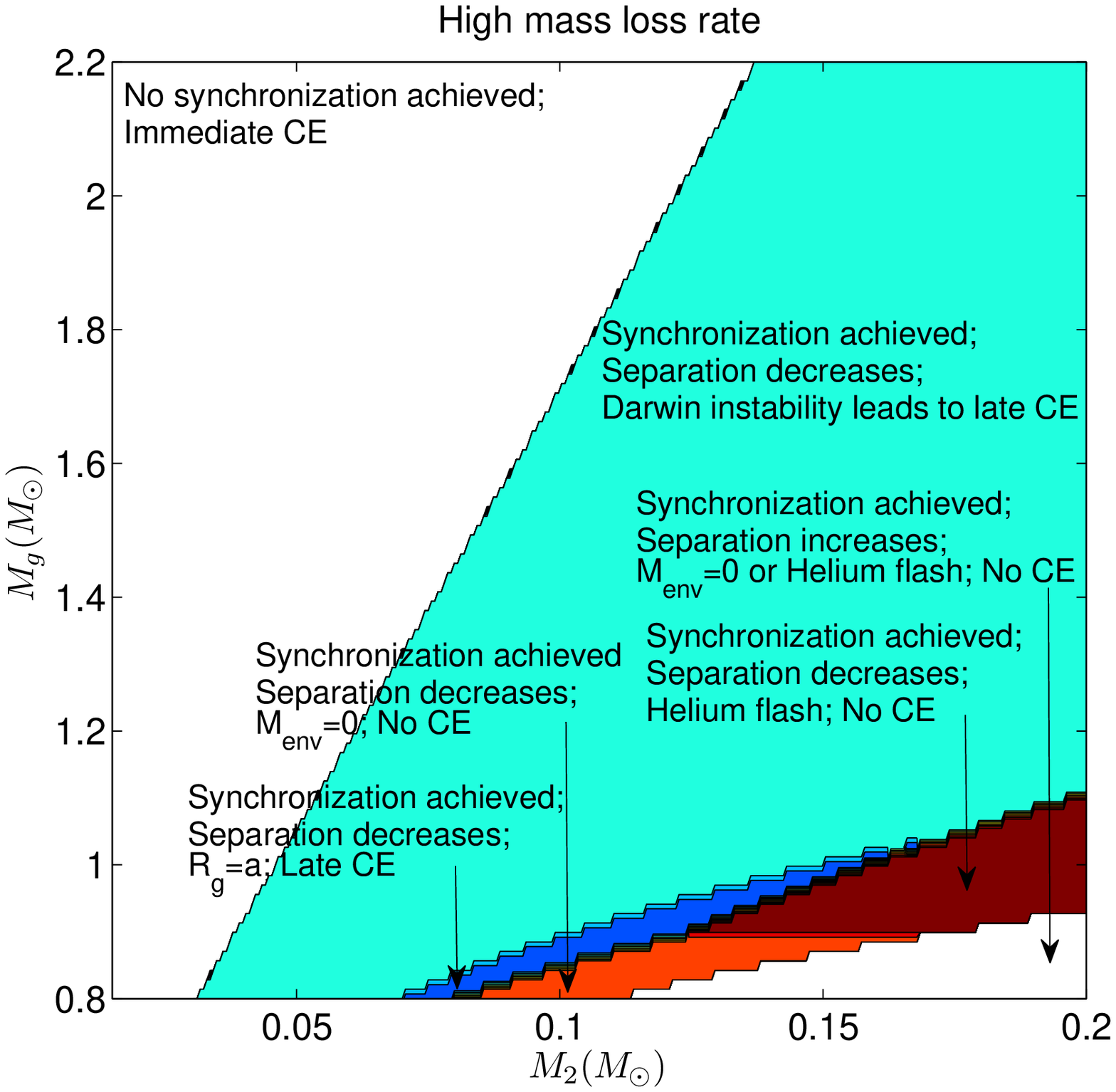}
\caption{Like Fig.~\ref{fig:sdbf3}, but for a high mass-loss rate
with
 $\eta_R=3 \cdot10^{-13}$. }
\label{fig:sdbf4}
 \end{figure}

We solved for the evolution according to the steps outlined in
section \ref{subsec:proceedure}. In Figs.~\ref{fig:sdbf3} and
~\ref{fig:sdbf4} we present the outcome of the evolution for our
two mass-loss rates: $\eta_R= 3 \times 10^{-14}$ (normal) and
$\eta_R=3 \times 10^{-13}$ (high), respectively, in equation
(\ref{dlx5}). In these sets of calculations the initial core mass
is $M_{c}(0)= 0.4M_\odot$, such that the RGB radius is $R_{g0}=
81R_\odot$. As strong tidal interaction starts when $a=5R_g$ by
our standard parameterizations, so that the initial binary
separation is $a= 405R_\odot$.

For the normal mass-loss rate (Fig. \ref{fig:sdbf3}) the envelope
mass decreases slowly, and so does its moment of inertia. As the
RGB radius increases, the Darwin instability sets in and the
companion spirals-in to the RGB envelope to form a CE. The CE
outcome covers a large area in the $M_2-M_g$ plane. It is
interesting to compare this case to the case where the RGB
evolution is not considered (Fig. \ref{fig:sdbf2}; Soker 2002).
When the increase in core mass and radius are not considered, only
a small parameters-area in the $M_2-M_g$ plane leads to the
formation of a CE phase after a stable synchronization has been
achieved. When the increase in radius is taken into account, this
area is much larger. For a small area in the $M_2-M_g$ plane
(lower-right) the core reaches a mass of $M_c=0.48 M_\odot$, for
which a core helium flash is assumed to occur (turning the star to
a horizontal branch star), and the binary system does not enter a
CE phase. Only much later during the AGB phase the system might
enter a CE phase.

For the high mass-loss rate (Fig. \ref{fig:sdbf4}) the envelope
mass decreases more rapidly, and so does its moment of inertia.
This makes the Darwing instability less likely. For that, in some
cases the RGB radius reaches the orbital separation before a
Darwin instability occurs. Also, the higher mass-loss rate leads
to the depletion of the envelope before a Darwin instability
occurs in some cases. In addition, mass loss acts to increase the
orbital separation. Therefore, for a larger range of
parameters-area in the $M_2 - M_g$ plane (lower-right corner) the
system does not enter a common envelope at all. Still, binaries
that located in a large area in the $M_2-M_g$ plane do enter a CE
phase after a stable synchronization has been achieved. For these
systems the RGB envelope mass decreases, in some cases
substantially, before a CE phase sets in.

\begin{figure}
\includegraphics[scale=0.6]{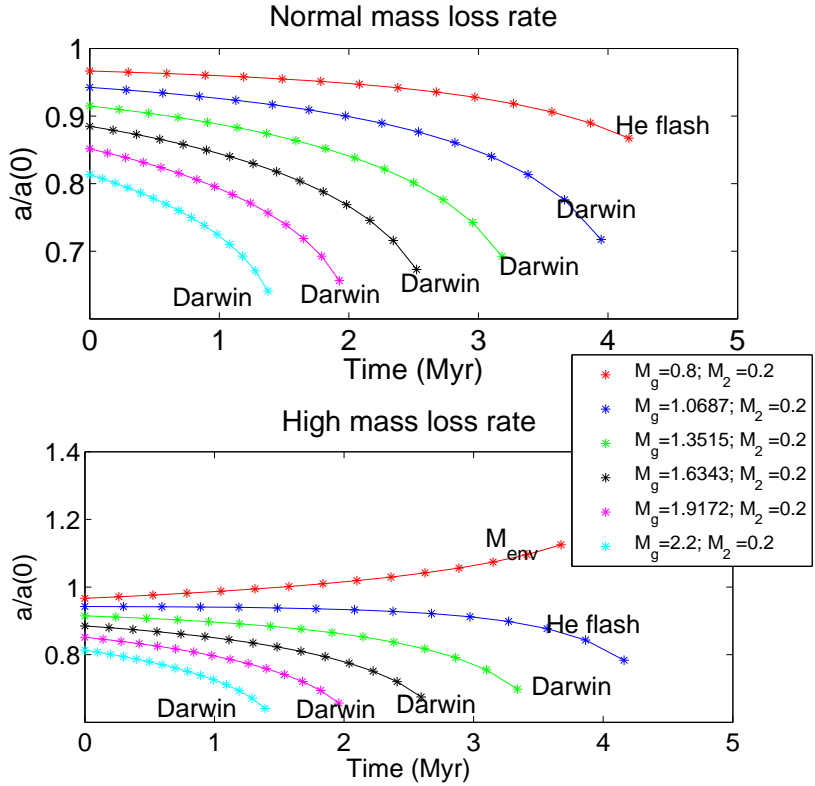}
\caption{Evolution of orbital separation in units of initial
orbital separation $a_0=5R_g(0)=405 R_\odot$, from the moment
synchronization is achieved until end of calculation, as explained
in the caption of Fig. \ref{fig:sdbf3}. Top panel is for the
normal mass-loss rate with $\eta_R= 3 \times 10^{-14}$ in equation
(\ref{dlx5}), and the bottom panel is for the high mass-loss rate
with $\eta_R=3 \times 10^{-13}$. The initial mass of the RGB star
is indicated for each run. All calculations use a companion mass
of $M_2=0.2 M_\odot$, and the RGB core mass at the beginning of
the calculation is as in the other runs in this section. }
\label{fig:sdbf5}
\end{figure}

\begin{figure}
\includegraphics[scale=0.6]{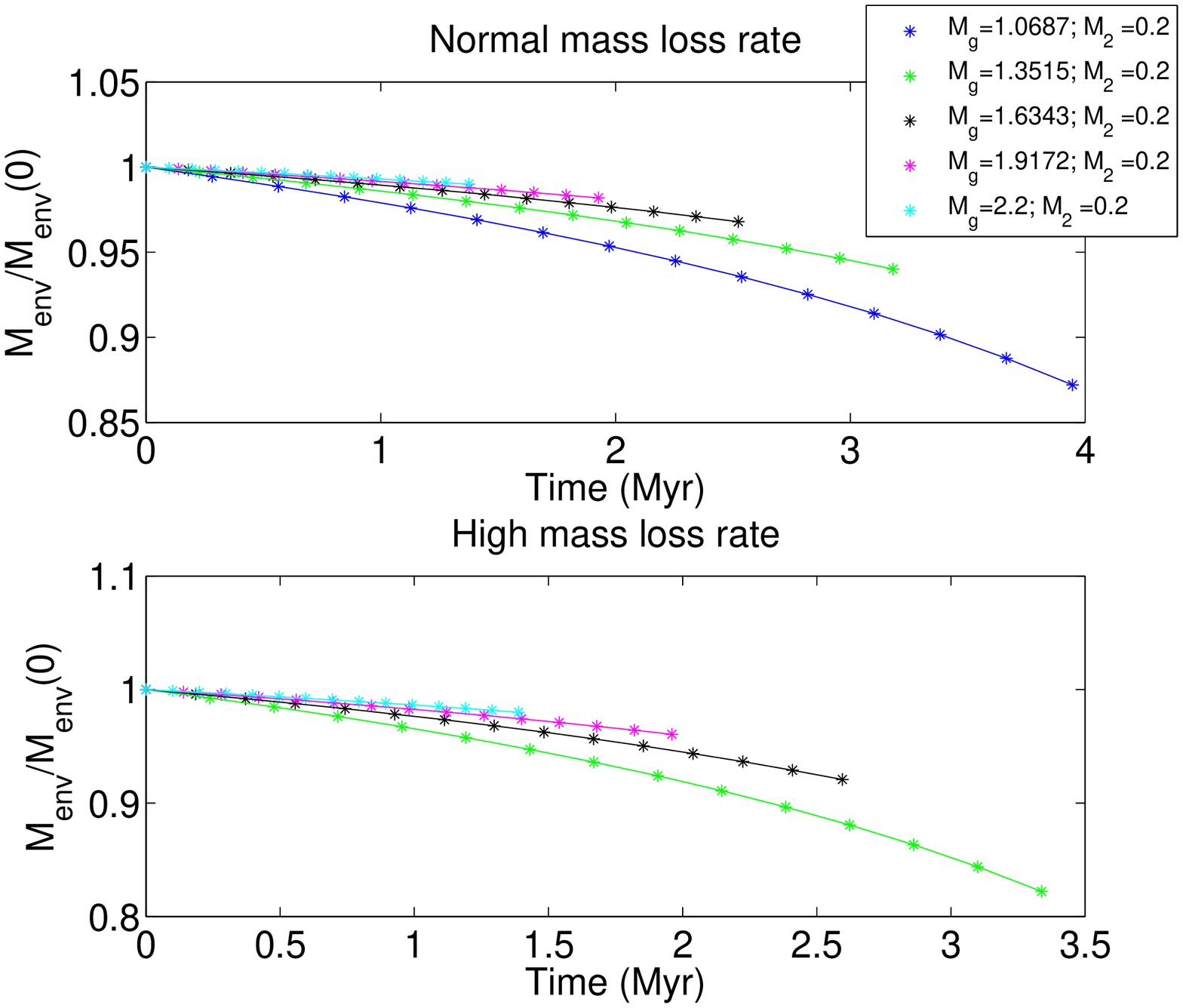}
\caption{Evolution of the envelope mass for the cases presented in
Fig. \ref{fig:sdbf5} that enter the CE phase. The envelope mass
$M_{env}$ is scaled to its initial value at the moment of
synchronization $M_{\rm env}(0)$. }
\label{fig:sdbf6}
 \end{figure}

In Fig. \ref{fig:sdbf5} we present the evolution of the scaled
orbital separation, $a/a_0$, with time, for the two values of the
mass-loss rate, and for several RGB initial masses. In all runs
$a_0= 405 R_\odot$ and $M_2=0.2M_\odot$, and all cases that are
shown reach a stable synchronized orbit after the initial tidal
interaction. The following characteristic of the evolution should
be noted.
\newline
(1) The values of $a/a_0$ start with values of $a/a_0 < 1$, because we assume that
the initial strong tidal interaction brings the system to synchronization in a short time.
\newline
(2) Since we start the calculation when the RGB is highly evolved (a core mass of
$M_{c}(0)=0.4 M_\odot$ and a radius of $R_{g0} = 81 R_\odot$), the evolution to the end of the
RGB is relatively short, $ \la 4 \times 10^6 \yr$.
If the initial orbital separation is smaller, then strong tidal interaction occurs earlier,
when the evolution is slower, and the evolution time will be longer.
\newline
(3) Synchronization implies that the RGB star rotates at about
$\ga 0.1$ times it break-up speed. Namely, $\Omega_g \ga 0.1$.
With its strong envelope convection, the synchronized RGB star is
likely to have strong magnetic activity. The rotation and magnetic
activity are likely to substantially enhance the mass-loss rate.
For that, we take the high mass-loss rate cases to represent our
expectation for the mass-loss rate.
\newline
(4) As expected, a higher mass-loss rate reduces the chances for
the formation of a CE. Still, by Fig. \ref{fig:sdbf4}, for a large
parameters-area a CE will occur after a substantial mass of the
envelope has been lost. The decreases of the envelope mass with
time until the onset of a CE phase is presented in Fig.
\ref{fig:sdbf6}. The envelope mass is scaled with its value at the
beginning of our calculation (when the system reaches
synchronization for the first time).

\subsubsection{Envelope rotation}
\label{subsec:rotation}

We saw that many systems form a CE phase only after the envelope
mass has been substantially reduced. Another important effect in
all the evolutionary routes considered here is the spinning-up of
the RGB-star envelope before the CE phase starts even. With their
strong convection, RGB stars that rotate even at a moderate rate
can have a relatively very high mass-loss rate.

The envelope angular velocity $\omega_g$ should be compared with the break up velocity
(the velocity of a test body performing a circular Keplerian orbit on the stellar equator)
\begin{equation}
\omega_{B} \equiv \sqrt{ \frac{G M_g}{R_g^{3}}}.
\label{dlx13}
\end{equation}
The important quantity here is the ratio $\Omega_g  \equiv
\omega_g/\omega_B$.

The orbital velocity of the RGB envelope $\omega_g$ changes during
the evolution. We will consider four points in the possible
evolutionary routes:
$\Omega_{g0}:$ for systems that did not reach a stable synchronized orbit, this is the
angular velocity of the envelope when the secondary just enters the envelope;
$\Omega_{g1}:$ when stable synchronization is first achieved;
$\Omega_{g2}:$ when a system with a stable synchronized orbit encounters the
Darwin instability and loses synchronization;
$\Omega_{g3};$ the primary angular velocity when a system that suffers a Darwin
instability just enters the CE phase.

\begin{figure}
\includegraphics[scale=0.6]{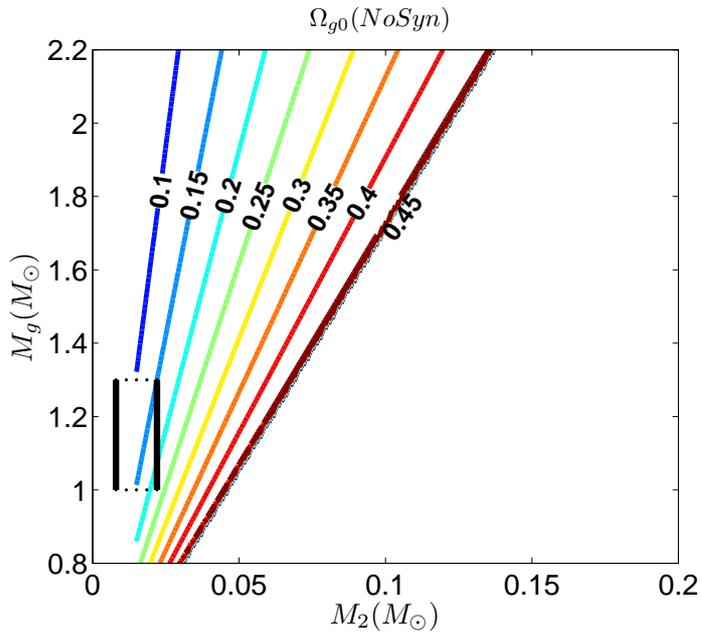}
\caption{The envelope angular velocity scaled by the break up
velocity when the companion enters the envelope ($\Omega_{g0}$).
The systems presented here did not achieve stable synchronized orbit at all,
and therefore the secondary spiralled in directly into the RGB
envelope. The black rectangle enclosed the possible area occupied
by the progenitor of HD149382 as we discussed in section \ref{sec:summary}. }
\label{fig:sdbf7}
\end{figure}

At the first time that tidal interaction takes place, either a
stable synchronization occurs or the companion rapidly enters a CE
phase. In the case where no stable synchronization occurs, we find
that when the companion just enters the RGB envelope the envelope
angular velocity is
\begin{equation}
\Omega_{g0} = \frac {\omega_g}{\omega_B} = \frac{\mu}{\eta
M_{\rm env}} \sqrt{ \frac{M_g+M_2}{M_g}} \left(
\sqrt{\frac{a_0}{R_{g0}}} -1 \right) \simeq 6\frac{M_2}{M_{\rm
env}},  \qquad {\rm no~synchronization}, \label{dlx14}
\end{equation}
where $\mu=M_g M_2/(M_g+M_2)$ is the reduced mass, and where in the last equality we
have substituted our standard values of $\eta=0.2$, $a_0=5 R_g$, and took $M_2 \ll M_g$.
We can substitute $M_{\rm env} = M_{g0}-0.4 M_\odot$, and recast equation (\ref{dlx14}) in the form
\begin{equation}
M_{g0}   \simeq 6 \frac{M_2}{\Omega_{g0} } + 0.4 M_\odot ,
\qquad {\rm no~synchronization}. \label{dlx14b}
\end{equation}
This is a straight line in the $M_2-M_g$ plane, as evident in Fig.
\ref{fig:sdbf7}, where we present the values of $\Omega_{g0}$ when
the companion enters the envelope for systems that did not reach
stable synchronization. It is evident from Fig. \ref{fig:sdbf7}
that even when the companion spirals-in very rapidly to the RGB
envelope, it manages to spin-up the primary to high rotation
velocity. Such rotation speeds are expected to enhance the
mass-loss rate, even before considering the gravitational energy
that is released by the secondary as it spirals-in inside the
envelope.

\begin{figure}
\includegraphics[scale=0.6]{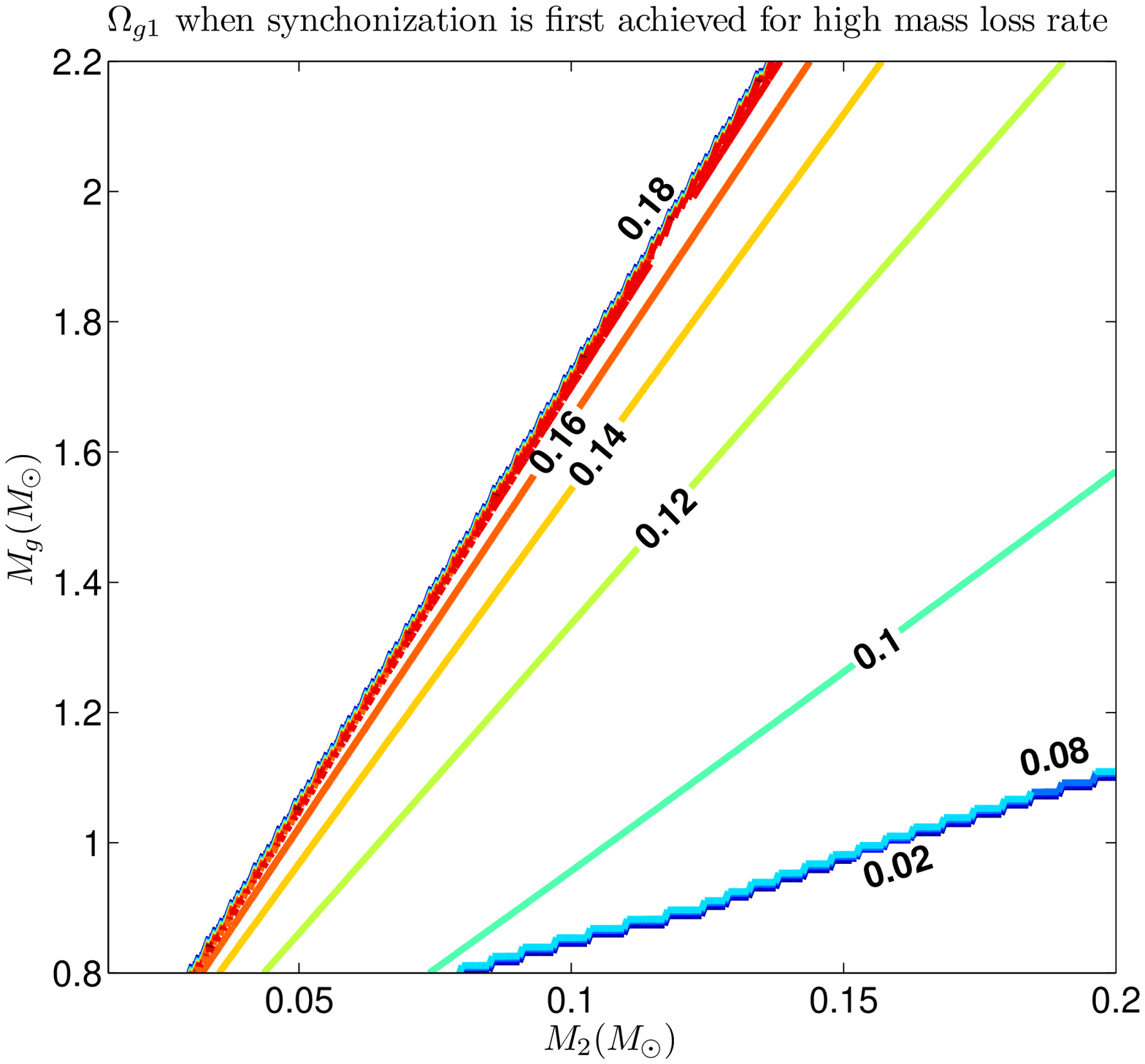}
\caption{The envelope angular velocity scaled by the break up
velocity when stable synchronization is first achieved
($\Omega_{g1}$). This figure has the same parameters as Fig.
\ref{fig:sdbf4}, but here only systems that reach synchronization
and  form a CE later in the evolution are shown. The mass-loss
parameter has $\eta_R = 3\times 10^{-13}$ in equation
(\ref{dlx5}). } \label{fig:sdbf8}
\end{figure}

Consider systems that do reach stable synchronized orbit. The
scaled envelope's angular velocity when synchronization is first
achieved, $\Omega_{g1}$, for these systems is presented in
Fig.~\ref{fig:sdbf8}. The condition to achieve a stable
synchronized orbit is given in equation (\ref{eq:darwin3}). (In
eq. \ref{eq:darwin3} the assumption $M_2 \ll M_g$ was used, while
in the numerical solutions presented here the exact equations are
used; differences are very small). When such a stable synchronized
orbit is achieved, the envelope angular velocity, by definition,
is equal that of the orbital angular velocity as given by
Eq.(\ref{dlx3}). It is evident that when synchronization is
achieved the primary rotates quite rapidly, and strong magnetic
activity and enhanced mass-loss rate are expected. Note that more
massive secondaries bring the RGB envelope to synchronization when
they are at larger orbital separations, and hence these systems
have lower values of $\Omega_{g1}$.

\begin{figure}
\includegraphics[scale=0.6]{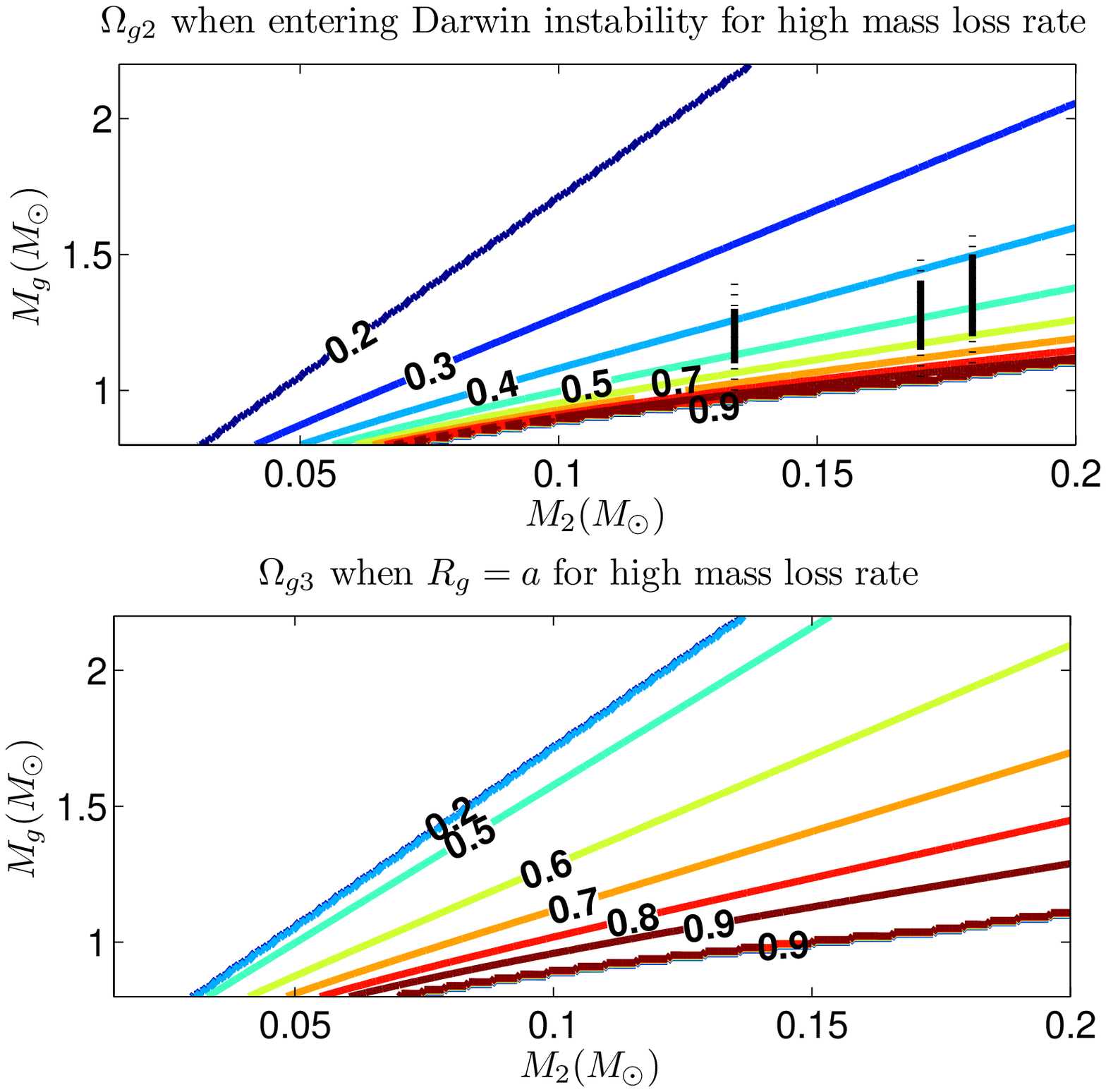}
\caption{The envelope angular velocity scaled by the break up
velocity. Top panel represents the envelope angular velocity when
the Darwin instability occurs ($\Omega_{g2}$).
Bottom panel represents the envelop`s angular
velocity when the secondary enters the envelope ($\Omega_{g3}$).
The three vertical lines represent the possible
locations of the progenitors of the three systems discussed in
section \ref{sec:summary}. }
\label{fig:sdbf9}
\end{figure}

\begin{figure}
\includegraphics[scale=0.6]{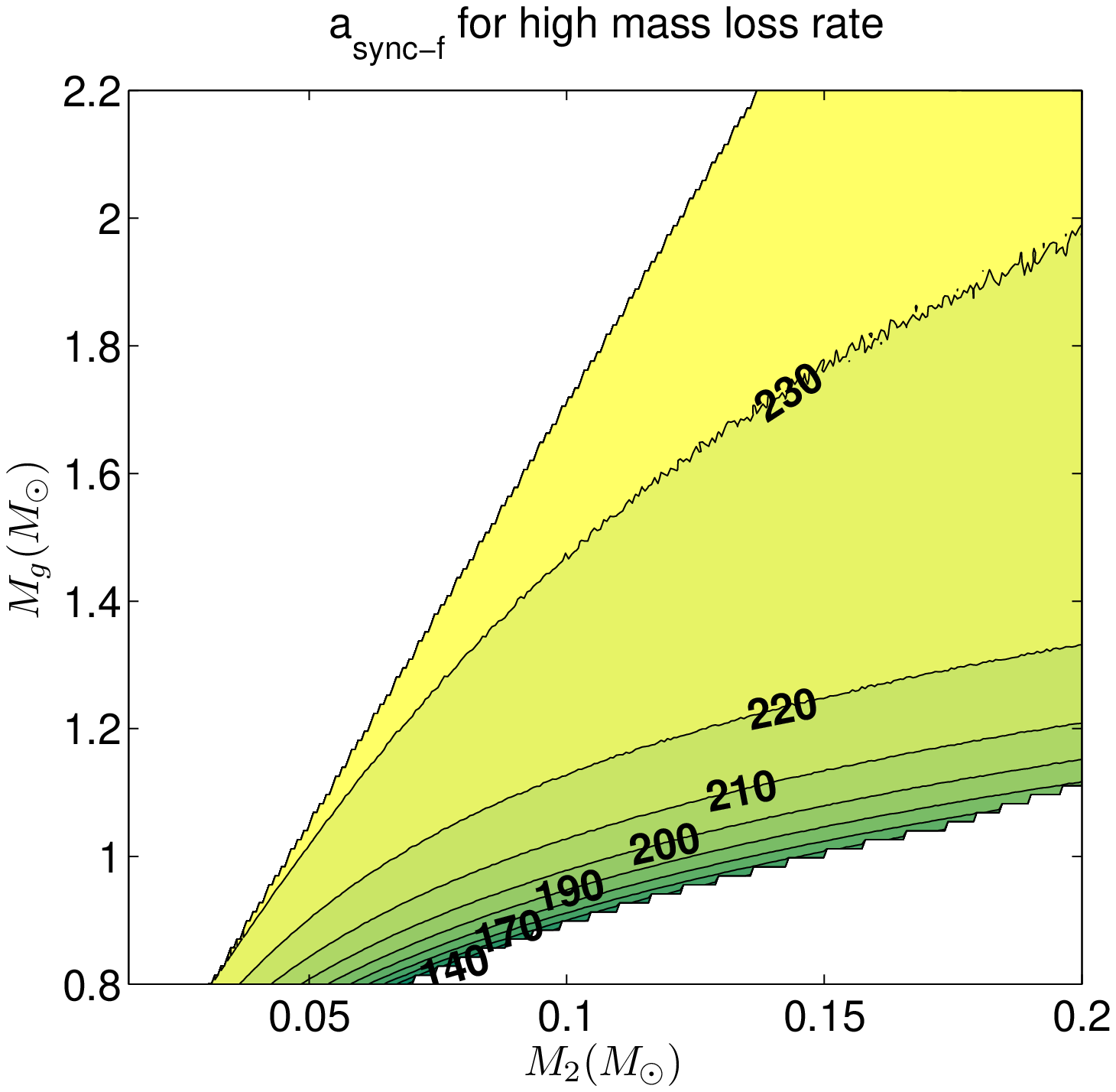}
\caption{Orbital separation when the Darwin instability occurs (and synchronization
is lost) for the same systems shown in Fig. \ref{fig:sdbf9}. }
\label{fig:sdbf10}
 \end{figure}

As can be seen in Fig.~\ref{fig:sdbf4}, most binary systems that have a stable synchronized
orbit encounter the Darwin instability, and the secondary rapidly spirals-in to
form a CE.
The normalized RGB angular velocity at the moment the system encounters the Darwin instability,
$\Omega_{g2}$, is presented in the top panel of Fig.~\ref{fig:sdbf9}.
The orbital separation at that moment is presented in Fig.~\ref{fig:sdbf10}.
As the secondary rapidly spirals-in during the instability mode, it transfers orbital
angular momentum to the envelope, further spinning it up.
The normalized RGB angular velocity at the moment the secondary of such a system
enters the envelope, $\Omega_{g3}$, is presented in the bottom panel of Fig.~\ref{fig:sdbf9}.
Fig.~\ref{fig:sdbf9} shows that the primary can reach very high angular velocity
prior to the formation of the CE, and further strengthens our point that the pre-CE evolution
must be considered.

In should be noted that although the end result of the systems presented
in Fig.~\ref{fig:sdbf7} and in Fig.~\ref{fig:sdbf9} is the formation of a common envelope,
there is a fundamental difference between
these two evolutionary routes.
The systems presented in Fig.~\ref{fig:sdbf7} form a CE shortly after a strong tidal
interaction takes place, while those presented in Fig.~\ref{fig:sdbf9} had evolved
through a stable synchronized orbit phase.
During that phase the RGB star can lose a substantial fraction of its envelope.

\section{DISCUSSION AND SUMMARY}
\label{sec:summary}

We have shown that for a large parameters space binary systems
reach stable synchronized orbit before the onset of a common
envelope (CE) phase. Although we have made the calculations for
RGB stars when their core mass at the beginning of the calculation
is $M_{c}(0)=0.4 M_\odot$, the results have wider implications.
Such stable orbits can be achieved for other evolved stars, like
AGB stars, and for evolutionary stages earlier or later than those
used here. Such stable orbits allow the RGB (or AGB, or other
evolved giant stars) to lose mass before the onset of the CE. Even
after the secondary enters the giant envelope, the rotation
velocity is high enough to cause enhanced mass-loss rate.

Our results imply that in many cases, studies that neglect the
pre-CE evolution (and start with a secondary on the surface of a
non-rotating giant) might reach some wrong conclusions. For
example, they might conclude that there is a need for extra energy
source to expel the envelope, because they do not consider the
mass-loss process prior to the onset of the CE phase.

Our results can be applied to specific systems.
We consider here four systems of sdB stars with companions:
\astrobj{HW Vir} with $(M_1, M_2, a)=(0.54 M_\odot, 0.18 M_\odot, 0.89 R_\odot)$
(Wood et al. 1993);
\astrobj{PG~1336-018} with $(M_1, M_2, a) =(0.50 M_\odot, 0.17 M_\odot, 0.79 R_\odot)$ (Kilkenny et al. 1998);
\astrobj{HS~0705+6700} with $(M_1, M_2, a)=(0.483M_\odot, 0.134 M_\odot, 0.81 R_\odot)$ (Drechsel et al. 2001);
and
\astrobj{HD149382} (Geier et al. 2009) with $(M_1, M_2, a)=(0.29-0.53 M_\odot, 0.008-0.022 M_\odot, 5-6.1 R_\odot)$
\newline

We mark the possible locations of the progenitors of the first
three systems on Fig. \ref{fig:sdbf9}. The initial mass of the
primary is not well constrained, but these systems most probably
went through a phase of a stable synchronized orbit. The mass-loss
rate could have been high, and a non-negligible amount of the RGB
envelope has been expelled before the formation of a CE. This
increases the survivability of the secondary. If a CE was formed
only after the RGB envelope expanded, this can explain the
observations that the core was massive enough to experience a core
helium flash after the onset of the CE.

The possible location of the progenitor of HD149382 in the
$M_g-M_2$ plane is marked on Fig. \ref{fig:sdbf7}. This system did
not reach synchronization, and a CE phase was formed immediately
after a strong tidal interaction occurred. However, the companion
did substantially spin-up the envelope, and by that increased the
mass-loss rate. This reduced the envelope mass, and by that
increased the survival probability of the companion.

This research was supported by the Asher Fund for Space Research
at the Technion, and the Israel Science foundation.

E.B. was supported in part by the Center for Absorption in
Science, Ministry of Immigrant Absorption, State of Israel.


\end{document}